\documentstyle[preprint,aps,epsfig]{revtex}
\draft
\begin{document}

\title{Ground state properties of ferromagnetic metal/conjugated
polymer interfaces}

\author{S.J. Xie$^{1,2}$,
K.H. Ahn$^1$, D.L. Smith$^1$, A.R. Bishop$^1$, and A. Saxena$^1$}
\address{$^1$Theoretical Division, Los Alamos National Laboratory,
Los Alamos, New Mexico 87545 \\$^2$School of Physics and Microelectronics,
Shandong University, Jinan, 250100, Peoples Republic of China}

\date{\today}

\maketitle

\begin{abstract}
We theoretically investigate the ground state properties of
ferromagnetic metal/conjugated polymer interfaces. The work is
partially motivated by recent experiments in which injection of
spin polarized electrons from ferromagnetic contacts into thin
films of conjugated polymers was reported.  We use a
one-dimensional nondegenerate Su-Schrieffer-Heeger (SSH)
Hamiltonian to describe the conjugated polymer and one-dimensional
tight-binding models to describe the ferromagnetic metal.  We
consider both a model for a conventional ferromagnetic metal, in
which there are no explicit structural degrees of freedom, and a
model for a half-metallic ferromagnetic colossal magnetoresistance
(CMR) oxide which has explicit structural degrees of freedom. The
Fermi energy of the magnetic metallic contact is adjusted to
control the degree of electron transfer into the polymer.  We
investigate electron charge and spin transfer from the
ferromagnetic metal to the organic polymer, and structural
relaxation near the interface. Bipolarons are the lowest energy
charge state in the bulk polymer for the nondegenerate SSH model
Hamiltonian.  As a result electrons (or holes) transferred into
the bulk of the polymer form spinless bipolarons. However, there
can be spin density in the polymer localized near the interface.
\end{abstract}

\pacs{PACS numbers: 72.25.-b, 73.61.Ph, 75.30.Vn, 71.38.-k}

\section{Introduction}
Magnetoelectronics or spintronics is a field of growing interest.
Since the discovery of giant magnetoresistance (GMR)
\cite{Baibich}, rapid progress has been made in this field.
Electron spin injection and spin dependent transport are essential
aspects of spintronics and have been extensively studied in a
number of different contexts including: ferromagnetic metals to
superconductors \cite{Meservey}; ferromagnetic metals to normal
metals \cite{Johnson}; ferromagnetic metals to nonmagnetic
semiconductors \cite{Monzon} and magnetic semiconductors to
nonmagnetic semiconductors \cite{Ohno}. Recently, spin polarized
injection and spin polarized transport in conjugated polymers have
been reported \cite{Dediu}. Specifically, spin injection was
observed into thin films of the conjugated organic material
sexithienyl from half-metallic manganites (in which electron spins
at the Fermi surface are completely polarized) at room
temperature. The ease of fabrication and low temperature
processing of conjugated organic materials open many application
possibilities, and electronic as well as opto-electronic devices
fabricated from these materials, e.g. organic light-emitting
diodes and spin valves, are being actively pursued \cite{Dediu}.

Theoretical study of spin polarized injection and transport has
been carried out primarily in the framework of classical transport
equations \cite{Son,Schmidt1,Zutic}. The role of interface
properties for spin injection in inorganic semiconductors was
investigated in this context
\cite{Rashba,Smith,Albrecht,Schmidt2}. The purpose of this paper
is to study the ground state characteristics such as lattice
displacements, charge density and spin density distribution of
conjugated organic polymers contacted with a ferromagnetic metal.
An added motivation to study this type of ``active'' interface is
that because of relatively large electron-phonon coupling the
materials on {\it both} sides of the interface can deform, which
may facilitate spin polarized injection.

The paper is organized as follows.  In the next section we present
tight-binding models for a nondegenerate conjugated polymer, a
ferromagnetic (FM) metal, a half-metallic colossal
magnetoresistance (CMR) material and the interface between the
polymer and the two kinds of magnetic materials.  Section III
presents the results for a model junction between the polymer and
the FM metal, and Sec. IV describes results for CMR/polymer
junctions. Our main findings are summarized in Sec. V.

\section{Model}
Organic polymers currently used for electronic and opto-electronic
devices typically have a nondegenerate ground state. The first
experimental evidence of spin polarized electrical injection and
transport in conjugated organic materials was carried out using
sexithienyl (T$_6$), a $\pi$-conjugated oligomer
\cite{Dediu,Muccini}. The underlying physics of spin injection and
transport is of particular interest for conjugated organic
materials, where strong electron-phonon coupling leads to
polaronic (or bipolaronic) charged states \cite{Waragai}. These
polymers or oligomers have chain structures that can be described
using a nondegenerate version of the one-dimensional SSH model,
the Brazovskii-Kirova (BK) model \cite{BK,Xie},
\begin{eqnarray}
H_{P}=-\sum_{i,\sigma}\epsilon_{P}a^+_{i,\sigma}a_{i,\sigma}
-\sum_{i,\sigma}[t_{P}-t_{1}(-1)^{i}-\alpha_{P}(u_{i+1}-u_{i})]
(a^+_{i,\sigma}a_{i+1,\sigma}+a^+_{i+1,\sigma}a_{i,\sigma})\nonumber\\
+\sum_{i}\frac{1}{2}K_{P}(u_{i+1}-u_{i})^2,
\end{eqnarray}
where $a^+_{i,\sigma}$ ($a_{i,\sigma}$) denotes the electron
creation (annihilation) operator at site $i$ with spin $\sigma$,
$\epsilon_{P}$ is the on-site electron energy of an atom, $t_{P}$
is the transfer integral in a uniform (undimerized) lattice and
$\alpha_{P}$ the electron-phonon interaction parameter, $t_{1}$
introduces nondegeneracy into the polymer, and $K_P$ denotes a
spring constant.

To describe a conventional ferromagnetic metal we use a 
one-dimensional tight-binding model with kinetic energy ($H_{ke}$) 
and spin splitting ($H_{Hund}$) terms:
\begin{eqnarray}
H_{FM} =H_{ke}+H_{Hund},
\end{eqnarray}

\begin{eqnarray}
H_{ke}=-\sum_{i,\sigma} t_{F}
       (a_{i,\sigma}^{+}a_{i+1,\sigma}+a_{i+1,\sigma}^{+}a_{i,\sigma}),
\end{eqnarray}

\begin{eqnarray}
H_{Hund}=-\sum_{i}J_{i}(a_{i,\uparrow}^{+}a_{i,\uparrow}-a_{i,\downarrow}^{+}
a_{i,\downarrow}),
\end{eqnarray}
where $t_F$ is the transfer integral for a ferromagnetic metal and
$H_{Hund}$ describes the spin splitting of the magnetic atom with
site-dependent strength $J_i$.  We take an occupation of one
electron per atom and $J_{i}=J_{M}$ with parameters $t_F=0.622$ eV
and $J_{M}=0.625$ eV for the conventional ferromagnetic metal.

CMR materials can form half-metallic ferromagnets and are
therefore very interesting as spin polarized electron injecting
contacts. CMR materials have a chemical composition such as
Re$_{1-x}$Ak$_{x}$MnO$_{3}$, where Re represents a rare earth
atom, e.g. La and Nd, and Ak represents an alkali metal such as
Ca, Sr and Ba. Mn has a valence of ($3+x$) which depends on the
doping concentration $x$. Depending on doping, the material can be
either a metal or an insulator and either ferromagnetic or
antiferromagnetic \cite{doping}.  In particular,
Re$_{1-x}$Ak$_{x}$MnO$_{3}$ can be a half-metallic ferromagnet
when $0.2<x<0.5$, for example, with Re=La and Ak=Ca.  In this state
all the electrons at the Fermi surface have the same spin
orientation.  The Mn atom has 5 electrons in its 3d orbitals.
These electrons have a parallel spin alignment due to a strong
Hund's rule splitting. Because of crystal field splitting in a
solid, three of the orbitals form the low energy $t_{2g}$ states,
$xy, yz$ and $zx$, and the other two form the higher energy
$e_{g}$ states, $x^{2}-y^{2}$ and $3z^{2}-r^{2}$. In the ground
state, the electrons in $t_{2g}$ are localized and constitute core
spins. The $e_{g}$ states are extended and the electrons in these
states can be delocalized. In cubic symmetry, the two $e_g$ levels
are degenerate. However in tetragonal or orthorhombic symmetry, the
degeneracy of $e_{g}$ is broken by a Jahn-Teller coupling due to
the electron-lattice interaction, which causes movement of oxygen
ions with respect to the manganese ions.

Here, we suppose for simplicity that a polymer or oligomer chain
is connected at the ends of a CMR lattice in the $z$-direction. We
consider a one-dimensional model which contains the basic
properties of a half-metallic CMR material; a ferromagnetic metal
with electron-lattice coupling. The following one-dimensional
model captures these essential features,
\begin{equation}
H_{CMR}=H_{ke}+H_{Hund}+H_{el-lat}+H_{elastic}, \\
\end{equation}
where $H_{ke}$ and $H_{Hund}$ are given in Eqs. (3) and (4), and
\begin{eqnarray}
H_{el-lat}=-\sum_{i,\sigma}\lambda_{F}(u_{i+1}-u_{i})a_{i,\sigma}^{+}a_{i,\sigma} ,
\\
H_{elastic}=\sum_{i}\frac{1}{2}K_{F}[(\delta_{i}-u_{i})^{2}+(u_{i+1}-\delta_{i})^{2}] .
\end{eqnarray}
Here $u_{i}$ and $\delta_{i}$ are the displacements of the ${i}$th
oxygen atom and manganese atom, respectively. $H_{ke}$ describes
electron hopping between two nearest manganese atoms. $H_{Hund}$
describes the spin splitting of a magnetic manganese atom that
results from interaction with the core spins. We have
$J_{i}=J_{M}$ for the ferromagnetic state (core spins aligned) and
$J_{i}=(-1)^{i}J_{M}$ for the antiferromagnetic state.
$H_{el-lat}$ gives the on-site energy of the manganese atoms,
which depends on the displacement of the nearest neighbor oxygen
atoms, and $\lambda_F$ denotes the electron-lattice coupling
strength. The last term $H_{elastic}$ represents the elastic
energy and includes nearest neighbor interactions.

Coupling at the interface between the conjugated polymer and the
ferromagnetic metal is described by the hopping integral,
\begin{equation}
t_{F-P}=\beta(t_F+t_P)/2 ,
\end{equation}
where $\beta$ is a weighting parameter. In principle, this
coupling could be spin-dependent, but here we take
$t_{F-P}^{\uparrow} =t_{F-P}^{\downarrow}=t_{F-P}$ for simplicity.
Periodic boundary conditions are adopted.

The parameters used for the CMR contact
Re$_{1-x}$Ak$_{x}$MnO$_{3}$ are $t_F=0.622$ eV, $J_{M}=1.25$ eV,
$K_{F}=7.4$ eV/{\AA}$^2$ \cite{Ahn} and $\lambda_{F}=2.0$eV/{\AA}.
For the organic polymer we take representative parameters as
$t_P=2.5$ eV, $\alpha_P=4.2$ eV/{\AA}, $K_P=21.0$ eV/{\AA}$^2$
\cite{Heeger}. We set the degeneracy breaking parameter $t_1$=0.04
eV so that the energy difference is $\epsilon_{AB}$=0.035 eV per
carbon atom between the two dimerized phases. The relative
chemical potential $\epsilon_P$ was used to adjust the electron
transfer between the ferromagnet and the polymer. Segment lengths
were taken so that Re$_{1-x}$Ak$_{x}$MnO$_{3}$ consists of 100 MnO
units and the polymer 100 of CH units, that is $N_{M}=N_{P}$=100.
For the most part, the results do not depend on the lengths of the
segments if they are not too short (i.e., much longer than the
characteristic polaron size). The interfacial coupling parameter
was taken as $\beta=1$. If $\beta>1$, the interface acts as a
potential well and tends to confine electrons, whereas if
$\beta<1$ the interface acts as a potential barrier and tends to
exclude electrons.

We first study an isolated Re$_{1-x}$Ak$_{x}$MnO$_{3}$ chain to
test the effectiveness of this model for the CMR material. The
electronic eigenstates
\begin{eqnarray}
|\psi_{\mu \sigma}> = \sum_{i} Z_{i,\mu,\sigma}
a_{i,\sigma}^{+}|0>
\end{eqnarray}
corresponding to the eigenvalue ${\varepsilon}_{\mu,\sigma}$ are
solved from the equation,
 \begin{eqnarray}
 -t_{F}Z_{i+1,\mu,\sigma}-t_{F}Z_{i-1,\mu,\sigma}
 -J_{i}\sigma Z_{i,\mu,\sigma}-\lambda_{F}(u_{i+1}-u_{i})Z_{i,\mu,\sigma}
 ={\varepsilon}_{\mu,\sigma}Z_{i,\mu,\sigma},
 \end{eqnarray}
where $\sigma=+1$ for spin up and $-1$ for spin down. The
displacements $\{\delta_i\}$ and $\{u_i\}$ in the ground state are
determined from the eigenstates self-consistently:
\begin{eqnarray}
\delta_{i}=\frac12(u_{i}+u_{i+1}) ,
\end{eqnarray}
\begin{eqnarray}
u_{i}=\frac12[\delta_{i-1}+\delta_{i}-\frac{\lambda_{F}}{K_{F}}
{\sum_{\mu,\sigma}}
(Z_{i,\mu,\sigma}Z_{i,\mu,\sigma}-Z_{i-1,\mu,\sigma}Z_{i-1,\mu,\sigma})].
\end{eqnarray}
If $\lambda_{F}=0$, the stable configuration has a uniform
structure, {\it i.e.} $\delta_i=0$ and $u_{i}=0$, without
distortion. Otherwise, some distortion will occur. From Eqs. (11)
and (12) we see that the displacements of both oxygen and
manganese atoms depend on the electronic density at the manganese
atoms.

The structure and magnetism of Re$_{1-x}$Ak$_{x}$MnO$_{3}$ depend
on the doping concentration $x$ which determines the electron
number per manganese atom. The orbitals of each manganese atom
have been renormalized to a single orbital in the present model
and the electron number per manganese atom is denoted by $y$ ($\le
1$). Figure 1($a$) shows the dependence of the energy difference
per site between the ferromagnetic (FM) and antiferromagnetic
(AFM) states on $y$. For an electronic doping concentration $y=0$
(no electrons), the FM and AFM states have the same energy and the
equilibrium conditions give $\delta_i=u_i=0$. For $y=1$, that is
each manganese atom having one electron, the AFM state is lower in
energy than the FM state. An energy gap of 2.5 eV appears in the
AFM state for both the spin up and down energy levels. The lower
subband levels are occupied and the system is an insulator.  At
$y=0.5$, the FM state is lower in energy than the AFM state. At
this electron concentration, the energy difference between FM and
AFM states is 0.127 eV per manganese atom. In this case, as shown
in Fig. 1($b$), the energy bands of the FM state are totally spin
split. There is a gap of 0.26 eV at the wavevector $k=\pi/2a$ ($a$
is the lattice constant between two nearest Mn sites) and the
system is an insulator. This gap can be adjusted by changing
$\lambda_F$. When $\lambda_F\le 1.4$ eV/\AA, the gap is close to
zero. All the spin down levels are empty and only the lower
sublevels of the spin up band are occupied. At $y=0.5$, the charge
density has an oscillatory distribution. For example, at
$\lambda_F=2.0$ eV/{\AA}, the densities on two adjacent manganese
atoms are about 0.621$e$ and 0.379$e$ (at $y=0.5$). Away from
$y=0.5$, the energy gap disappears and the system becomes a
ferromagnetic half-metal. In the ferromagnetic state, the sites
displace in the approximate pattern,
 \begin{eqnarray}
 \delta_{i}=\delta_{0}\sin[2i(y-0.5)\pi],\\
 u_{i}=u_{0}\cos[2i(y-0.5)\pi].
 \end{eqnarray}
With electron concentration $0.2\le y \le 0.45$, the displacements
of both Mn and O atoms become very small ($\delta_{0}\le 0.005$
\AA~ and $u_{0}\le 0.01$ \AA) and decrease to zero when the chain
length becomes arbitrarily long. That is, the system becomes
uniform in this doping region, and correspondingly the charge
density is also uniform with a half-metallic property.  These
results are consistent with the basic properties of the CMR
materials \cite{doping} and show that the one-dimensional model
gives a reasonable description of them.

\section{Ferromagnetic metal/polymer junction}
We first consider a polymer chain contacted by a simple rigid
ferromagnetic metal chain. In the case of half filling for the
parameters used, the spin polarization is
$\rho=(N_{\uparrow}-N_{\downarrow})/(N_{\uparrow}+N_{\downarrow})=
0.34.$  The polymer has a one-dimensional chain structure with a
strong electron-lattice interaction that will cause localized
charged excitations. When the polymer is connected with a
ferromagnetic metal, both the lattice configuration and charge
distribution of the polymer are affected.  By adjusting the
relative chemical potential, electrons (or holes) are transferred
into the polymer and cause the displacement of the lattice sites.
Results are shown in Fig. 2 and Fig. 3 for $\epsilon_P=0$ and
$\epsilon_P=1.0$ eV, respectively. Following the usual convention,
in this paper the displacement is plotted with a multiplying
factor $(-1)^i$, where $i$ is the site index. The ferromagnetic
metal is to the left and the polymer is to the right of the
interface which is between sites $n$=100 and 101. In Fig. 2, there
is no electron transfer between the segments, and the charge
density is uniform within the whole system. But the charges near
the interface can be spin polarized. The polarization oscillates
and decays into the polymer segment. The decay length of the spin
polarization is about 6$b$, where $b$ is the lattice constant in
the polymer.  If the chemical potential of the ferromagnet is
higher than the bipolaron level in the polymer, as in Fig. 3,
electrons are transferred into the polymer segment to reach a new
equilibrium for the system. Instead of forming extended electronic
waves, the extra electrons in the polymer form localized charged
bipolarons. Figure 3($a$) shows the displacements of lattice
sites, from which we see that, in this case, three complete
bipolarons are formed within the polymer together with some local
distortions at the interface. The corresponding charge and spin
distributions are shown in Fig. 3($b$). In the present BK model
for the nondegenerate polymer, bipolarons are energetically lower
than polarons. Since each bipolaron has two confined electronic
charges with opposite spins, a bipolaron has no spin. There is
neither localized nor extended spin distribution within the
polymer layer. Because the polymer is nonmagnetic in the ground
state, or more generally at thermal equilibrium, there is no spin
distribution far from the interface.

\section{Ferromagnetic CMR/polymer junction}
Here, we consider the polymer chain in contact with a
half-metallic ferromagnetic Re$_{1-x}$Ak$_{x}$MnO$_{3}$ chain with
electron concentration $y=0.32$. By adjusting the relative
chemical potential, electrons are transferred between the CMR
material and polymer. At $\epsilon_{P}=2.15$ eV, there is
essentially no electron transfer between the segments. Figure
4($a$) shows the displacements of the atoms (Mn, O and C) compared
to their uniform bulk positions. Within the CMR segment, both the
manganese and oxygen atoms are only slightly displaced. The small
displacements are due to the finite length of the segment and
disappear as the segment length is increased. The carbon atoms
have a displacement of 0.05 {\AA}, corresponding to the bulk
dimerization of the polymer chain. The interfacial atoms have a
deviation from the bulk dimerization, which results in a small
expansion of the end bonds of the CMR segment and a contraction of
the first few polymer bonds. The charge and spin densities are
shown in Fig. 4($b$). Because the CMR material is completely spin
polarized. the charge and spin densities coincide in this segment.
The distributions of charge and spin density in each segment are
uniform except for a small modulation near the interface. The
modulation in the CMR material is a finite size effect discussed
previously. There is neither a net charge nor spin distribution
within the bulk polymer. When we increase the chemical potential
of the CMR material, electrons are transferred into the polymer.
The results for $\epsilon_{P}=2.90$ eV are shown in Fig. 5. At
this value for the chemical potential, 6.11 electrons transfer to
the polymer segment. The CMR segment keeps a nearly uniform
lattice structure except a small deviation at the interface. In
the polymer, from the displacements shown in Fig. 5($a$),
bipolaron states form. The localized electronic density can be
seen more clearly in Fig. 5($b$). Because the transferred
electrons form spinless bipolarons, there is no spin amplitude
within the polymer segment.

These results become more apparent if we examine the change in
electronic density of states (DOS) defined by the Lorentz line 
shape formula
\begin{equation}
g_\sigma(\varepsilon)=\sum_{\mu}\frac{1}{\pi}\frac{\lambda}{(\varepsilon-
\varepsilon_{\mu \sigma})^2+\lambda^2} ,
\end{equation}
where $\varepsilon_{\mu \sigma}$ is a one-electron energy
eigenvalue and $\lambda$ a phenomenological Lorentz line width,
which we choose as $\lambda$=0.15 eV. Figure 6($a$) shows the DOS
for the CMR/polymer chain before coupling of the two segments
(that is, for the two separate material segments) and Fig. 6($b$)
shows the DOS after coupling. The relative chemical potential was
adjusted to be $\epsilon_{P}=2.15$ eV as in Fig. 4 so that there
is no electron transfer between the CMR material and polymer after
coupling. From the figure we see that there is still a large gap
for the spin down states, but the gap for spin up states decreases
after coupling. All the occupied states near the Fermi level have
spin up and these states are confined in the segment of the CMR
material. Increasing the relative chemical potential
$\epsilon_{P}=2.90$ eV as in Fig. 5, we plot the DOS before and
after coupling in Fig. 7($a$) and Fig. 7($b$), respectively.
Because the Fermi level of the CMR is above the bipolaron level of
the polymer, electrons transfer to the polymer after coupling.
They form double-charged bipolarons. The bipolaron levels are
indicated in Fig. 7($b$), where the levels of spin up and down
states overlap (the spin up states of the bipolaron near $-$2.5 eV
cannot be seen easily due to the large DOS caused by the CMR
material). Thus, bipolarons have no spin. The difference in DOS of
spin up and down at the bipolaron states arises from the effect of
the CMR material at the interface.

\section{conclusions}
Organic ($\pi$-conjugated) polymers differ from traditional
inorganic semiconductors due to their strong electron-lattice
interactions. Carriers in (nondegenerate) polymers are not
typically electrons or holes but rather charged polarons or
bipolarons. In this paper we have studied the ground state
properties of a ferromagnetic metal/organic polymer junction.  Two
kinds of magnetic contacts were considered, a simple ferromagnetic
metal with fractional spin polarization and a CMR ferromagnet with
a half-metallic ground state. It was found that the electric
charges in the polymer near the interface can have spin
polarization. The spin density decays in an oscillatory fashion
into the polymer. By adjusting the relative chemical potential,
electrons can be transferred into the polymer from the magnetic
layer through the interfacial coupling. The transferred electrons
form bipolarons, which have no spin, so that there is no spin
density in the bulk of the polymer.  The polymer is nonmagnetic
and in the ground state, or more generally at thermal equilibrium,
there will not be a spin polarization in this material far from
the interface.

The present model is simple and only static characteristics were
investigated. In addition, the model is one-dimensional and
Coulomb interactions between electrons were not included so that
there was no Schottky effect at the interface.  But major factors
have been included, such as lattice relaxation and interfacial
coupling. The main motivation for this model came from recent
polarized spin injection (and spin-coherent transfer) experiments
on the conjugated organic oligomer sexithienyl (thin film) in
which a half-metallic ferromagnetic CMR contact was used
\cite{Dediu}. This oligomer can serve as an active transport
material for potential organic opto-electronic and spintronic
devices. Dynamics under external bias will be studied to describe
spin injection and polarized spin transport in conjugated organic
materials, but an understanding of ground state properties is
required to initiate a study of such dynamics.

\section{Acknowledgment}
This work was supported by the SPINs program of the Defense
Advanced Research Projects Agency and in part by the U.S.
Department of Energy.

\begin{figure}[t]
\psfig{figure=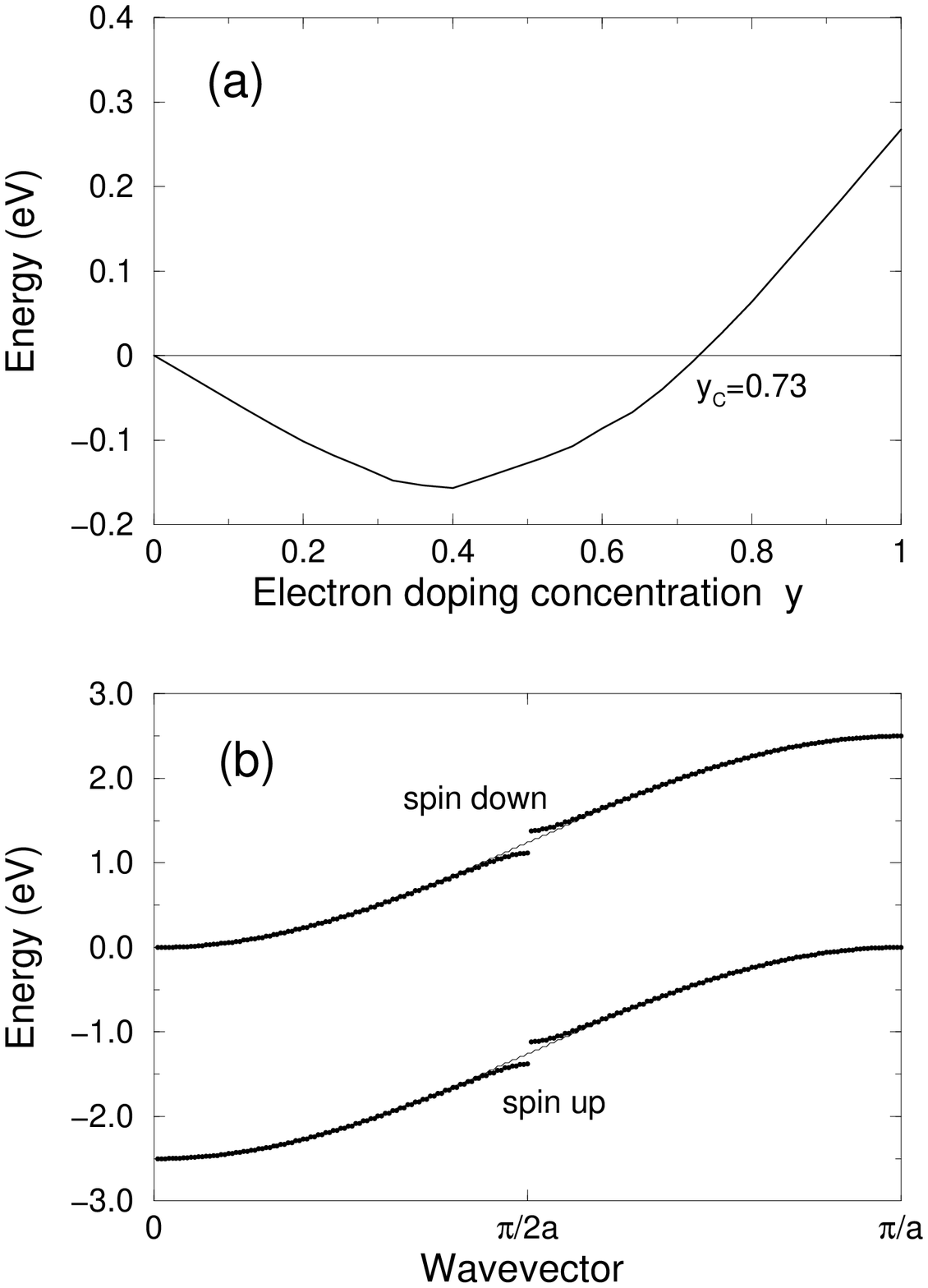,height=16.5cm,width=11.0cm,angle=0}
\caption[Fig1] { (a) Dependence of energy difference per site
between a one-dimensional ferromagnetic chain and an
antiferromagnetic Re$_{1-x}$Ak$_x$MnO chain with doping
concentration $y$. (b) Energy levels of Re$_{1-x}$Ak$_x$MnO in the
ferromagnetic state: $y=0.5$ (thick line) and $y=0.32$ (thin
line). The upper curve in panel (b) is for spin down electrons and
the lower curve is for spin up electrons.  A gap of 0.26 eV
appears at $k=\pi/2a$ in the case of half doping ($y=0.5$). }
\label{Fig1}
\end{figure}

\begin{figure}[t]
\psfig{figure=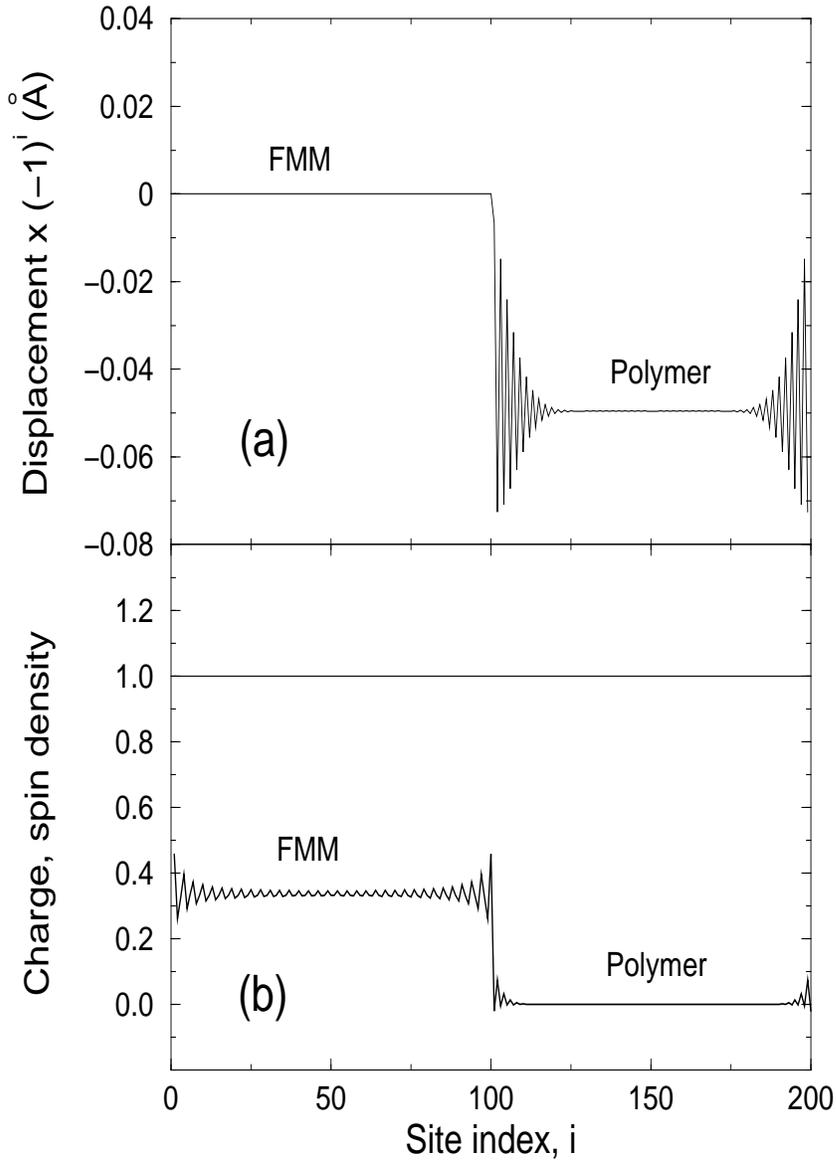,height=16.5cm,width=11.0cm,angle=0}
\caption[Fig2] { For a simple ferromagnetic metal (FMM)/polymer
chain: (a) Site displacements; (b) charge (thin line) and spin
(thick line) density distributions.  There is no electron transfer
between the FMM and the polymer. The interface is between sites
100 and 101 and $\epsilon_P=0$. All the site displacements are
plotted after multiplying with a factor $(-1)^i$, where $i$ is the
site index. } \label{Fig2}
\end{figure}

\begin{figure}[t]
\psfig{figure=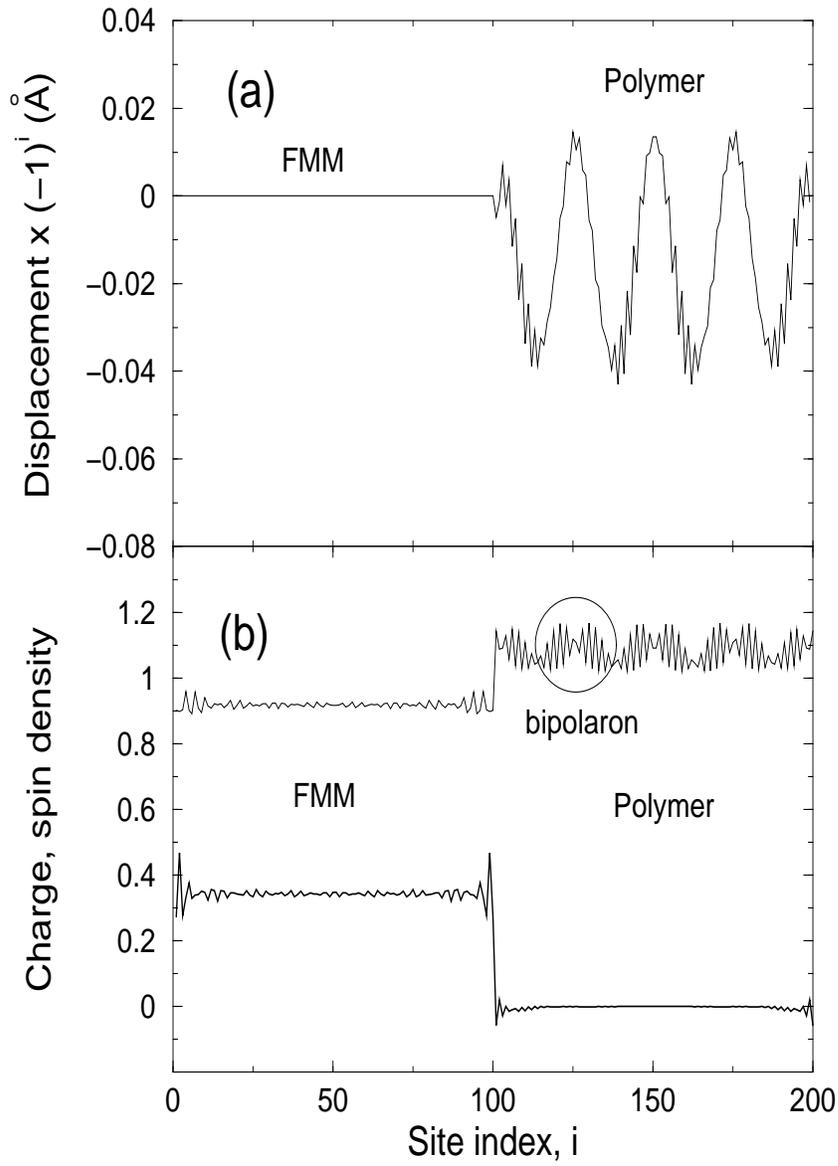,height=16.5cm,width=11.0cm,angle=0}
\caption[Fig3] { Same as in Fig. 2 but for $\epsilon_P=1.0$ eV.
Electrons are transferred from the FMM to the polymer through the
interface by increasing the chemical potential of the FMM,
resulting in bipolarons forming in the polymer. } \label{Fig3}
\end{figure}

\begin{figure}[t]
\psfig{figure=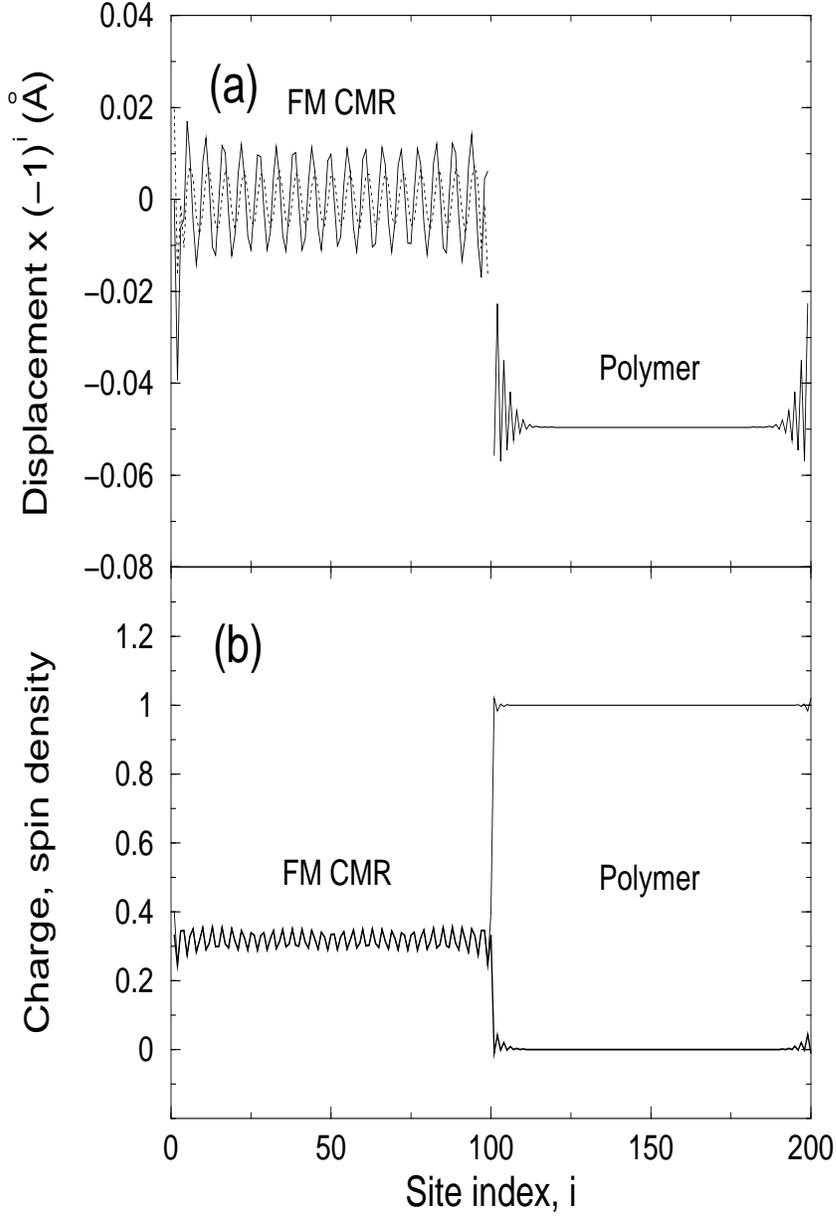,height=16.5cm,width=11.0cm,angle=0}
\caption[Fig4] { For a ferromagnetic Re$_{1-x}$Ak$_x$MnO (FM
CMR)/polymer chain: (a) Site displacements of Mn (left dotted), O
(left solid) and C (right solid) atoms; (b) charge (thin line) and
spin (thick line) density distributions. The charge and spin
densities coincide in the CMR material. There is no electron
transfer between the FMM and the polymer. The interface is between
sites 100 and 101.
 }
\label{Fig4}
\end{figure}

\begin{figure}[t]
\psfig{figure=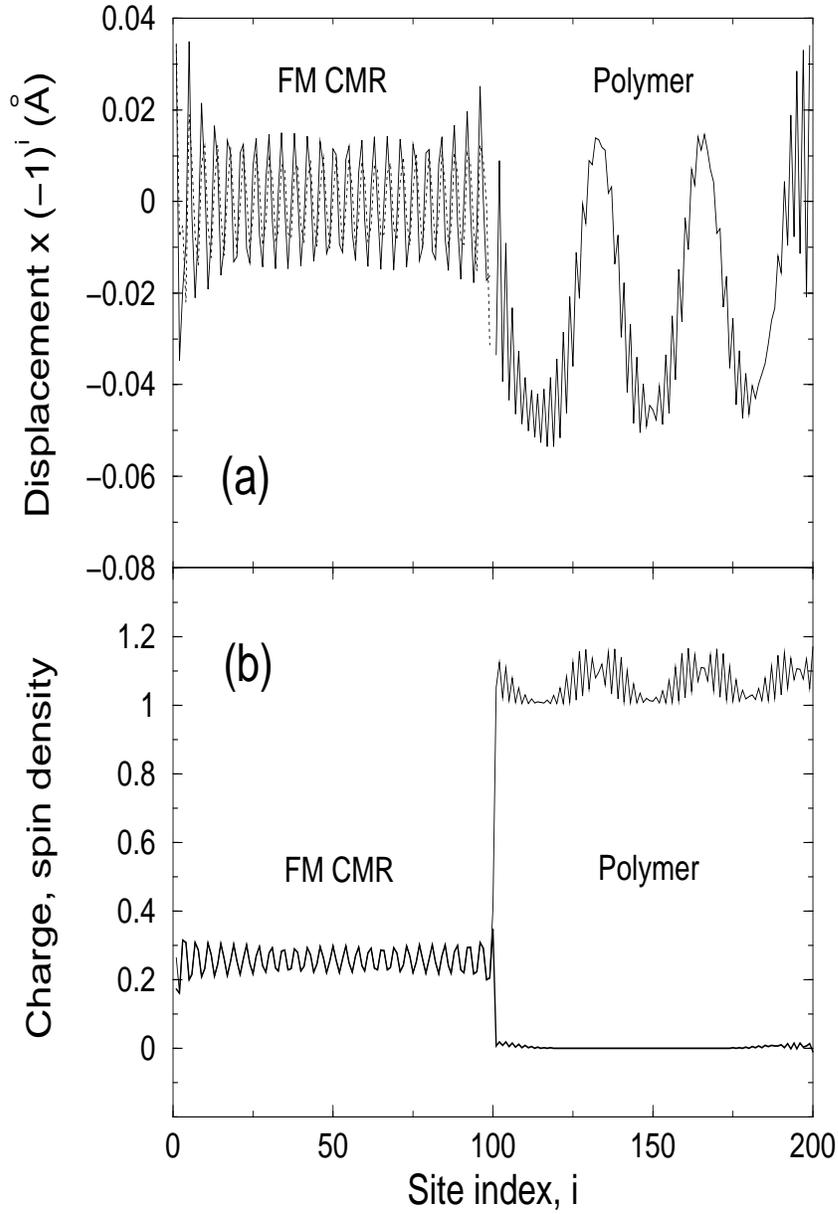,height=16.5cm,width=11.0cm,angle=0}
\caption[Fig5] { Same as in Fig. 4 but with some electrons
transferred from the FM CMR segment to the polymer through the
interface by increasing the chemical potential of the FM CMR
material, resulting in bipolarons forming in the polymer. }
\label{Fig5}
\end{figure}

\begin{figure}[t]
\psfig{figure=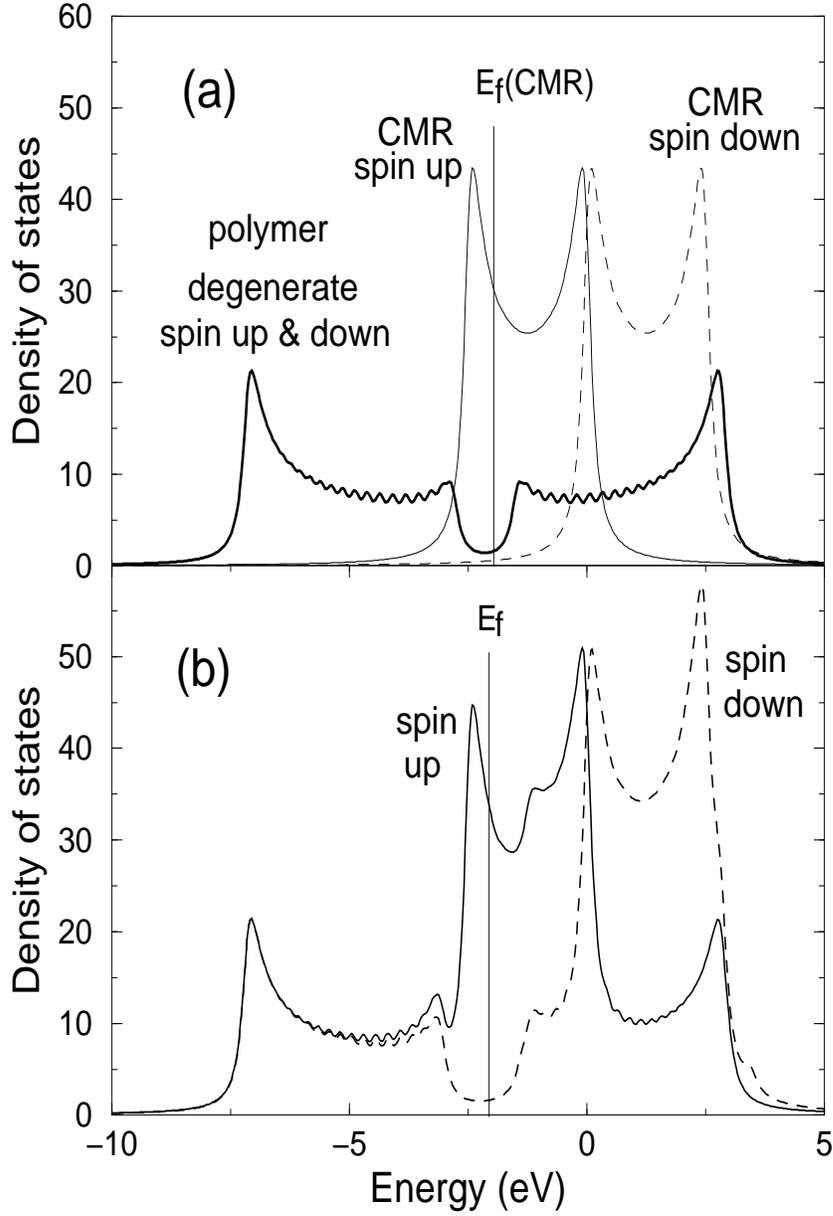,height=16.5cm,width=11.0cm,angle=0}
\caption[Fig6] { Density of states of the FM CMR material and the
polymer: (a) before coupling and (b) after coupling. The thick
solid line in (a) is for both spin up and spin down electrons in
the polymer, the thin solid (dashed) line in (a) is for spin up
(spin down) electrons in the polymer.   The solid (dashed) line in
(b) is for spin up (spin down) electrons. The phenomenological
Lorentz width is $\lambda$=0.15. The Fermi level of the CMR
material lies below the bipolaron energy of the polymer, so that
there is no significant electron transfer. } \label{Fig6}
\end{figure}

\begin{figure}[t]
\psfig{figure=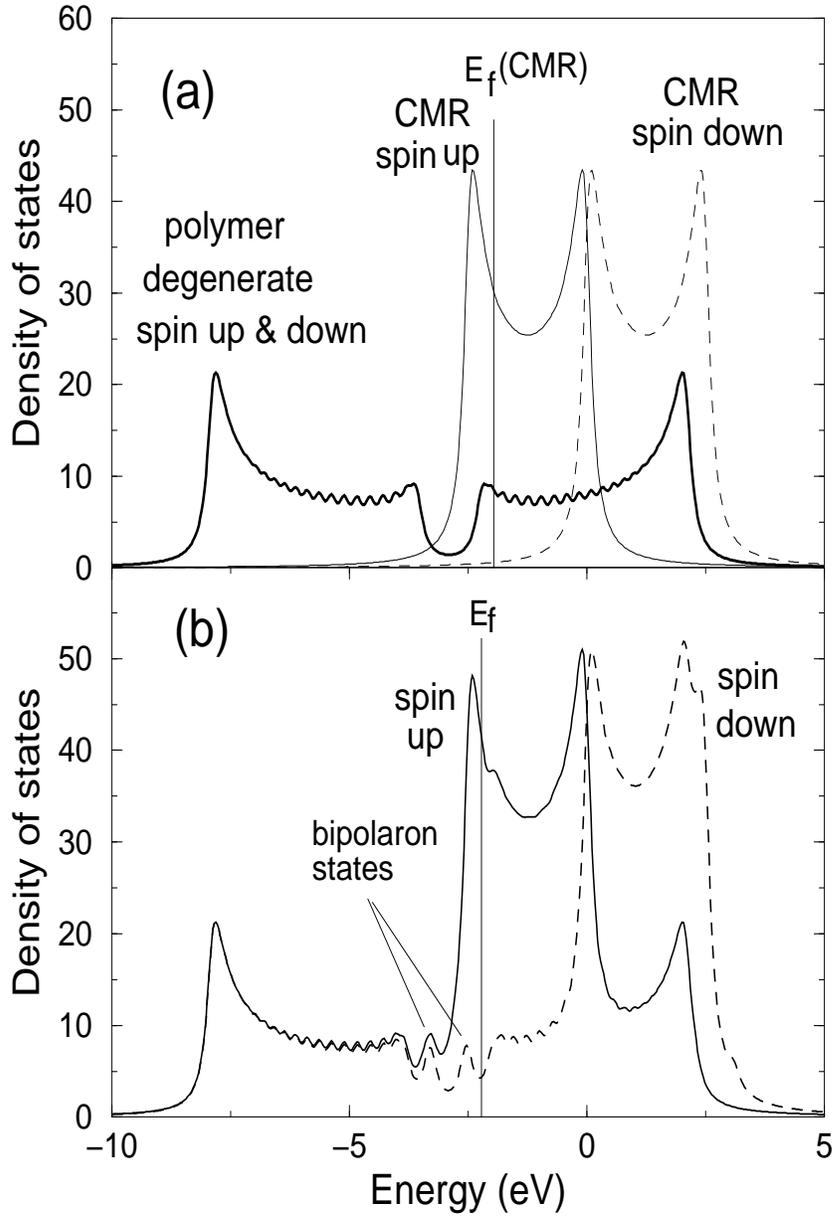,height=16.5cm,width=11.0cm,angle=0}
\caption[Fig7] { Same as in Fig. 6 but with the Fermi level of the
CMR material higher than the bipolaron energy of the polymer, so
that electrons transfer from the CMR segment to the polymer after
coupling. } \label{Fig7}
\end{figure}

\end{document}